\documentclass[final,5p,times,twocolumn,sort&compress]{elsarticle}

\usepackage[T1]{fontenc}
\usepackage{hyperref}

\usepackage{amsmath}
\usepackage{amsthm}
\usepackage{bm}
\usepackage{mathtools}
\usepackage{caption}
\usepackage{slashed}
\usepackage{cuted}
\usepackage[dvipsnames]{xcolor}
\usepackage{ulem}
\usepackage{subcaption}

\def\O{\mathcal O}
\def\MS{{\overline{\mathrm{MS}}}}
\def\RI{\mathrm{RI}}

\def\tr{\mathrm{tr}}

\def\Zg{Z^\gamma}
\def\Zp{Z^{\slashed p}}

\def\SMOM{\mathrm{SMOM}}

\journal{Physics Letters B}

\begin{document}

\begin{frontmatter}

\title{A note on momentum subtraction schemes for quark bilinears and semileptonic operators}

\author[e]{P.~A.~Boyle}
\author[a,b]{M.~Bruno\corref{cor1}}
\cortext[cor1]{Corresponding author}
\author[d]{M.~Gorbahn}
\author[g]{S.~J{\"a}ger}
\author[c]{C.~Lehner}
\author[f]{F.~Moretti}
\author[c]{J.~Parrino}

\affiliation[e]{
    organization={Physics Department, Brookhaven National Laboratory},
    addressline={Upton, NY 11973},
    city={Upton},
    state={NY},
    postcode={11973},
    country={USA}
}

\affiliation[a]{
    organization={Dipartimento di Fisica ``Giuseppe Occhialini'', Università degli Studi di Milano-Bicocca},
    addressline={Piazza della Scienza 3},
    postcode={20126},
    city={Milan},
    country={Italy}
}

\affiliation[b]{
    organization={Istituto Nazionale di Fisica Nucleare (INFN), Sezione di Milano-Bicocca},
    addressline={Piazza della Scienza 3},
    postcode={20126},
    city={Milan},
    country={Italy}
}

\affiliation[d]{
    organization={Department of Mathematical Sciences, University of Liverpool},
    postcode={L69 3BZ},
    city={Liverpool},
    country={United Kingdom}
}

\affiliation[g]{
    organization={Department of Physics and Astronomy
University of Sussex},
    postcode={Brighton BN1 9QH},
    city={Falmer},
    country={United Kingdom}
}

\affiliation[f]{
    organization={Institut f{\"u}r Theoretische Teilchenphysik, 
    Karlsruher Institut f{\"u}r Technologie},
    addressline={Wolfgang-Gaede Stra\ss{}e 1},
    postcode={76131},
    city={Karlsruhe},
    country={Germany}
}

\affiliation[c]{
    organization={Universität Regensburg, Fakultät für Physik},
    addressline={Universitätsstraße 31},
    postcode={93040},
    city={Regensburg},
    country={Germany}
}

\begin{abstract}
In this work we examine a family of regularization invariant (RI) symmetric momentum (SMOM) schemes for semi-leptonic operators. By working with chirally symmetric and massless QCD, we relate the semi-leptonic operators with their corresponding flavor-changing vector currents, whose renormalization in pure QCD is protected by the Ward identity. For the latter, we extend the original RI/SMOM scheme~\cite{Sturm:2009kb} to a family projectors suitable to be promoted to the semi-leptonic case and demonstrate their equivalence to Ref.~\cite{Gorbahn:2022rgl}, relevant in particular for the perturbative calculation of the corresponding Wilson coefficients.
\end{abstract}

\end{frontmatter}

\section{Introduction}
Due to the continuously growing precision of predictions from Lattice QCD simulations, choices of renormalization schemes matter alongside the inclusion of isospin-breaking corrections, which introduces additional complications (from the definition of the photon field in a finite box to possible difficulties in the analytic continuation to Minkowski signature). 
In this manuscript we review the construction of renormalized operators suitable for lattice QCD calculations. More specifically, we focus on semi-leptonic operators, relevant in the study of hadronic $\tau$ decays or in $\pi_{\ell 2}$, $\pi_{\ell 3}$, $K_{\ell 2}$ and $K_{\ell 3}$ transitions, to name a few, all phenomenologically extremely relevant (for recent examples see for instance Refs.~\cite{Bruno:2018ono,ETMCtau24,Boyle:2022lsi,DiCarlo:2019thl,DiCarlo:2019knp}). The state-of-the-art framework used for such predictions employs the weak Effective Field Theory (EFT), defined by the product of a Wilson coefficient and the corresponding flavor-changing four-fermion operator
\begin{equation} \label{eq:operatordef}
    \mathcal O_{rs}(x) = (\bar \nu_\ell \gamma^\mu P_L \ell)(x) (\overline \psi_r \gamma_\mu P_L \psi_s)(x) \,,
\end{equation}
with $P_L = \frac{1-\gamma_5}{2}$ and $\psi_r$ quark fields of flavor $r$. Note that we do not specify the family of the neutrino-lepton pair, since our results apply to several channels. Although the operator in Eq.~(\ref{eq:operatordef}) is schematically written as a product of two currents, the renormalized operator cannot be taken to be the product of the renormalized currents,  beyond the isospin limit $m_u = m_d$ and $e = 0$, and a renormalization prescription for the four-fermion operator must be specified.

When the operator $\O_{rs}(x)$ is inserted inside a given matrix element and calculated using Lattice QCD simulations one must adopt renormalization schemes suitable for this framework.
In particular the $\MS$ scheme based on dimensional regularization cannot be (directly) used without passing first through a different scheme\footnote{See also Ref.~\cite{Bruno:2017iwk} for a discussion on the matching between theories with and without a weak bosons with non-perturbative methods.}, and it is limited to perturbation theory by construction.
A class of schemes that can be implemented both on the lattice and in perturbation theory (with dimensional regularization) is given by the regularization invariant momentum (RI-MOM) subtraction schemes~\cite{Martinelli:1994ty,Sturm:2009kb,Chetyrkin:1999pq,Gracey:2003yr,Aoki:2007xm,Boyle:2016wis,Aoki:2005ga}, which have witnessed substantial progress in the determination of the Wilson coefficient for the operator $\O_{rs}(x)$ \cite{Gorbahn:2022rgl}, where a perturbative calculation achieves the change of scheme to the $\MS$ in which the Wilson coefficient and its running is easily obtained. Other possibilities commonly used in Lattice QCD calculations are the Schr\"odinger Functional renormalization scheme~\cite{Luscher:1991wu,Luscher:1993gh,deDivitiis:1994yz,Jansen:1995ck,Sint:2010eh}, position space schemes~\cite{Martinelli:1997zc,Gimenez:2004me} and the more recent Gradient flow~\cite{Luscher:2010iy,Luscher:2011bx,Luscher:2014kea}. 

The leading logarithmic radiative short-distance corrections have been calculated long ago (in the context of hadronic $\tau$ decays) in the so-called W-regularization scheme \cite{Marciano:1988vm,Sirlin:1977sv,Sirlin:1981ie} (see also Refs. \cite{Braaten:1990ef,Erler:2002mv}). 
Recently, the Wilson coefficient for the operator $\O_{rs}$ has been completed up to $O(\alpha \alpha_s)$, both in the $\MS$ scheme and in several RI variants \cite{Brod:2008ss,Gorbahn:2022rgl,Cirigliano:2023fnz}. In this work, we discuss generalized RI renormalization conditions suitable for the semi-leptonic operator, and, more specifically, we show how to express the schemes introduced in Ref.~\cite{Gorbahn:2022rgl} in terms of a compact double-trace projector, which fits with the treatment of bilinear operators in Ref.~\cite{Sturm:2009kb}

For the short-distance corrections, we retain the exact dependence on the strong coupling while truncating isospin-breaking effects to leading order in $e$\footnote{An approximation sufficient with the current level of precision needed for example in hadronic $\tau$ decays for the muon anomalous moment \cite{Aliberti:2025beg,Bruno:2018ono}.} and neglecting the quark masses. 
Now, in the (pure-QCD) isospin limit, the quark current in Eq.~(\ref{eq:operatordef}) does not renormalize and decouples from the leptonic one, such that the semi-leptonic operator does not need to be renormalized either. For the quark current, the conservation is reflected through field-theoretic Ward identities which are respected by the RI-SMOM schemes defined in the literature \cite{Gracey:2011fb,Sturm:2009kb}.

The same is not true, however, for the definition of the RI schemes for the semi-leptonic operators as originally introduced in Ref.~\cite{Carrasco:2015xwa}. This then leads to a non-minimal renormalization of the operator in pure QCD. The origin of the renormalization prescription can be traced back to a single-trace convention used in the scheme definition, together with a choice of projector inherited from the four-quark operator case where the issue is not relevant. This situation was rectified in Ref.~\cite{Gorbahn:2022rgl} by deriving an appropriate single-trace projector resulting in RI schemes where the operator normalization is $1 + \mathcal{O}(\alpha)$, and residual (perturbative) uncertainties are significantly reduced. 

In the present work, we reformulate the schemes of Ref.~\cite{Gorbahn:2022rgl} by employing a double-trace prescription, which results in a more compact projector and makes for a clearer connection with the bilinear case. We stress that the schemes themselves are unchanged, and in particular all results of Ref.~\cite{Gorbahn:2022rgl} apply.

The remainder of the manuscript is organized as follows. In Section \ref{sec:framework}, we present the overall framework of our study and summarize basic relations among Green's functions. In Section \ref{sec:wf}, we define our conventions for the wave-function renormalization factors and in Section \ref{sec:left} we discuss the relation between the semi-leptonic and the vector bilinear operator. In Section \ref{sec:vector}, we present our main result with the derivation of projectors for which radiative corrections in the strong coupling (alone) are absent, before concluding.

\section{Theoretical framework}
\label{sec:framework}

We start by listing the working assumptions of our study. Firstly, we work with massless QCD, as is the standard choice for renormalization schemes like $\MS$ or $\RI$. 
Secondly, the schemes we consider are defined by conditions imposed on the renormalized, four-dimensional Green's functions, whereby we require that the Ward identities implied by isosymmetric QCD apply.  We therefore do not need to specify a regularization scheme -- the primary virtue of the RI schemes. In practice, we may imagine a lattice discretization of QCD retaining good chiral symmetry at finite cutoff and with a lattice spacing sufficiently fine to ignore discretization errors. These particular assumptions simplify the practicalities of carrying out the renormalization on the lattice, much like dimensional regularization simplifies the perturbative renormalization of gauge theories, but nothing in the following depends on them.

We begin our treatment by introducing the expectation value of the quark propagator for flavor $r$
\begin{equation}
    S_r(p) = \int d^4x \, e^{ipx} \, \langle \psi_r(x) \overline \psi_r(0) \rangle \,.
\end{equation}
All fields and operators are bare unless explicitly specified and expectation values are taken in isosymmetric QCD. In our derivation, we restrict ourselves to momenta sufficiently large to ignore the effects of spontaneous chiral-symmetry breaking~\cite{Politzer:1976tv,Pascual:1981jr,Blum:2001sr,Aoki:2007xm}. In this limit and in the absence of a mass term, the inverse propagator is~\footnote{Here we follow the convention of Ref.~\cite{Gorbahn:2022rgl} where $\Sigma(p^2)= 1 + O(\alpha_s)$.}
\begin{equation}
    S_r^{-1}(p) = -i \slashed p \, \Sigma_r(p^2) \,.
    \label{eq:Sinv}
\end{equation}
Next we focus on the two-point Green's function in momentum space for generic kinematics,
\begin{equation}
    \Gamma_{\mu,rs}^X(p,p') = \int d^4x \, d^4y \, e^{ipx - ip'y} \, \langle \psi_r(x) [\overline \psi_r \Gamma^X \psi_s](0) \overline \psi_s(y) \rangle \,,
    \label{eq:Gamma}
\end{equation}
evaluated with off-shell gauge-fixed external states for several Dirac structures,
\begin{equation}
    \Gamma^V = \gamma_\mu \,, \quad
    \Gamma^A = \gamma_\mu \gamma_5 \,, \quad
    \Gamma^L = \gamma_\mu P_L\,.
\end{equation}
Since we take massless quarks, once the Wick contractions in Eq.~\eqref{eq:Gamma} are performed and only the connected contribution survives for $r\neq s$, the flavor index can be dropped from the notation.
The amputated Green's functions
\begin{equation}
    \Lambda^X_\mu(p,p') = S^{-1}(p) \, \Gamma_\mu^X(p,p') \, S^{-1}(p') \,,
\end{equation}
obey the relations (for sufficiently large momenta)
\begin{align}
    \Lambda_\mu^A = \Lambda_\mu^V \gamma_5 \,, \quad \Lambda_\mu^L = \Lambda_\mu^V P_L \,.
    \label{eq:LambdaL_LambdaV}
\end{align}

In our notation, spin and color indices are open; for example the tree-level value for the vector amplitude $\gamma_\mu$ is understood as $[\gamma_\mu]^{\alpha\beta} \delta^{ab}$. In the rest of the manuscript, we will highlight the presence of colors with factors $N_c$ and we will denote with $\tr[\dots]$ the trace over both spin and color indices.

Momentum subtraction schemes are fully specified by the choice of the external states (i.e. by the momentum configuration and the choice of the QCD gauge) together with projectors in spin and color space. 
We impose $p^2 = p'^2= -\mu^2$ and we introduce
\begin{equation}
    q_\mu = (p-p')_\mu \,.
\end{equation}

\section{Wave function renormalization}
\label{sec:wf}

The fist step in our program is the renormalization of the fields, generically given by
\begin{equation}
    \psi^R = Z^{1/2} \, \psi \,.
\end{equation}
Since the subtraction point is uniquely defined, we drop from the notation the dependence on the scale $p^2=-\mu^2$.
In the RI/MOM scheme the wave-function renormalization factor, which we denote with $\Zg$, is fixed by the condition \cite{Martinelli:1994ty}
\begin{equation}
    \frac{1}{16 N_c} \tr\left[ i \frac{\partial S_R^{-1}(p)}{\partial p_\mu} \gamma_\mu \right] = 1 \,,
\end{equation}
which implies
\begin{equation}
    \Zg = \Sigma(p^2) + \frac{p^2}{2} \Sigma'(p^2) \,.
\end{equation}

Instead, we label with $\Zp$ the quark field renormalization factor in the RI'/MOM and RI/SMOM schemes, fixed by the different condition
\begin{equation}
    \frac{1}{4 N_c p^2} \tr\left[i \slashed p \, S_R^{-1}(p^2) \right] = 1 \,,
\end{equation}
which, using Eq.~\eqref{eq:Sinv}, leads to the relation
\begin{equation}
    \Zp = \Sigma(p^2) \,.
    \label{eq:Zp}
\end{equation}
The two different wave-function renormalization factors can be easily related to each other according to
\begin{equation}
    \Zg - \Zp = \frac{p^2}{2} \Sigma'(p^2) \,.
    \label{eq:Zg_Zp}
\end{equation}

\section{Four-fermion operator}
\label{sec:left}

For a generic wave-function renormalization factor $Z$, renormalized amputated amplitudes read
\begin{equation}
    \Lambda_\mu^{X,R}(p,p') = Z_X Z^{-1} \Lambda_\mu^{X}(p,p') \,.
\end{equation}
By introducing a generic projector $[P_\mu]^{\alpha\beta,ab}$ obeying
\begin{equation}
    \tr[P^\mu \gamma_\mu] = 1 \,,
\end{equation}
we impose the following renormalization condition on the vector flavor-changing bilinear operator
\begin{equation}
    Z_V \frac{\tr[P^\mu \Lambda_\mu^V(p,p')]}{Z} = 1 \,.
    \label{eq:ZV}
\end{equation}

The next step in our discussion is the amputated four-fermion amplitude $\Lambda_\O$. Since our goal is to focus on QCD corrections, the four-fermion amplitude factorizes in the isosymmetric limit according to
\begin{equation}
\begin{split}
    [\Lambda_\O(p,p')]^{\alpha\beta\gamma\delta,ab} & \, = [\Lambda_\mu^L(p,p')]^{\alpha\beta,ab} [\gamma^\mu P_L]^{\gamma\delta} \\ & \, \equiv [\Lambda_\mu^L(p,p')] \otimes [\gamma^\mu P_L] \,.
\end{split}
\end{equation}

Renormalization conditions of the form
\begin{equation}
    \mathcal P[\Lambda_\O(p,p')] = 1 \,,
\end{equation}
may be imposed both in pure QCD and subsequently in the full theory, using a projector with the general form
\begin{equation}
    P^{\alpha\beta\gamma\delta,ab} = [P^\mu P_R]^{\alpha\beta,ab} [ \gamma_\mu P_R]^{\gamma\delta} = [P^\mu P_R] \otimes [\gamma_\mu P_R] \,,
    \label{eq:proj}
\end{equation}

with the understanding that its action is specified by
\begin{equation}
    \mathcal P [ \Lambda_\mathcal O(p,p') ] \equiv  P^{\beta\alpha\delta\gamma,ba} [\Lambda_\mathcal O (p,p')]_{\alpha\beta\gamma\delta,ab}  \,.
\end{equation}
Notice how the equation above defines a double-trace scheme differing from Ref.~\cite{Gorbahn:2022rgl} based on a single trace.
In the absence of isospin-breaking effects, the renormalization of the left current and of the semi-leptonic operator is completely fixed by its vector component
\begin{equation}
    \mathcal P[\Lambda_{\mathcal O}(p,p')] = \tr[P^\mu \Lambda_{\mu}^V( p, p')] \,.
\end{equation}
In fact, thanks to chiral symmetry and by demanding the additional property $\{P_\mu, \gamma_5\} = 0$, we have (for $p,p',q \gg \Lambda_\mathrm{QCD}$)
\begin{equation}
    \tr[P^\mu \gamma_5 \Lambda^A_\mu(p,p')] = -\tr[P^\mu \Lambda_\mu^V(p,p')] \,,
\end{equation}
which implies that
\begin{equation}
    \tr[P^\mu P_R \Lambda_\mu^L(p,p') ] = \frac12 \tr[P^\mu \Lambda_\mu^V(p,p')] \,,
\end{equation}
namely the renormalization of the axial and left currents, in QCD, is completely fixed by $Z_V$ which becomes the central investigation of our study.

The bare amputated amplitude of the vector operator for general kinematics obeys the Ward Identity
\begin{equation}
    q^\mu \Lambda_\mu^V(p,p') = i S^{-1}(p) - iS^{-1}(p') \,.
    \label{eq:WI}
\end{equation}
Once expressed in terms of renormalized fields and operators, the Ward identity,
\begin{equation}
    q^\mu \Lambda_\mu^{V,R}(p,p') = i Z_V \left( S_R^{-1}(p) - S_R^{-1}(p') \right) \,,
\end{equation}
keeps its functional form if $Z_V=1$. As a consequence the renormalization condition in Eq.~\eqref{eq:ZV} turns into
\begin{equation}
    \tr[P^\mu \Lambda_\mu^V(p,p')] = Z(p^2)\,.
    \label{eq:ref}
\end{equation}
When this is the case, once the projector and operator are promoted to the semi-leptonic case pure QCD corrections are absent order by order, making them preferable in general, but also for the perturbative calculation of the corresponding Wilson coefficients~\cite{Gorbahn:2022rgl}. Therefore in the rest of the manuscript, we focus on schemes in which we can prove that $Z_V=1$, or said differently, where we have equations of the form \eqref{eq:ref}.

\section{Vector current}
\label{sec:vector}

We restrict the discussion to the popular MOM and SMOM kinematic configurations, defined by $q_\mu=0$ and $q^2=-\mu^2$ respectively. 
Since in both cases $p^2$, $p'^2$ and $p \cdot p'$ are all linearly dependent, a generic tensor decomposition of the amplitude is
\begin{equation}
    \Lambda_\mu^{V}(p,p') = \gamma^\nu T_{\mu\nu}^i(p,p') F_i(p^2) \,,
    \label{eq:tensor}
\end{equation}
with $F_i(p^2)$ scalar functions.

\subsection{Exceptional kinematics}

In the original RI/MOM proposal the focus was on exceptional kinematics~\cite{Martinelli:1994ty}. Despite being less appealing in high-precision modern calculations, e.g. due to the pion-pole contamination at low momenta \cite{Politzer:1976tv,Pascual:1981jr,Martinelli:1994ty,Blum:2001sr,Aoki:2007xm}, we discuss it to illustrate (non-perturbatively) why the choice of projector is important. This will set the stage for the more-involved non-exceptional kinematics, the target of this work.

Here, the Ward identity in Eq.~\eqref{eq:WI} reduces to
\begin{equation}
    \Lambda_\mu^V(p,p) = i \frac{\partial S^{-1}(p)}{\partial p^\mu} \,.
    \label{eq:WI2}
\end{equation}
By examining the derivative of the propagator on the r.h.s., we immediately notice that the relevant Lorentz structures are $\gamma_\mu$ and $p_\mu \slashed p$, fixing the tensor in Eq.~\eqref{eq:tensor} to be~\cite{Gracey:2003yr}
\begin{equation}
    T_{\mu\nu}^i(p,p) = \left\{ g_{\mu\nu} \,, \frac{p_\mu p_\nu}{\mu^2} \right\}^i \,,
    \label{eq:Tmunu_MOM}
\end{equation}
with $g_{\mu\nu}$ denoting the Minkowski metric.
By saturating Eq.~\eqref{eq:WI2} with $\gamma^\mu$, we find
\begin{equation}
    \frac{1}{16 N_c} \tr\left[ \gamma^\mu \Lambda_\mu^V(p,p) \right] = \Zg \,,
\end{equation}
which automatically implies that $Z_V=1$ when the amputated Green's function is renormalized with $\Zg$. Instead, when we project the amplitude with the other structure, via eqs.~\eqref{eq:Zp} and \eqref{eq:Zg_Zp} we find
\begin{equation}
    \frac{1}{4 N_c p^2} \tr[\Lambda_\mu^V p^\mu \slashed p] = 4 \Zg - 3 \Zp \,.
\end{equation}
By defining the appropriate linear combination that cancels the term proportional to $\Zg$, we recover the projector~\cite{Gracey:2003yr}
\begin{equation}
    P_\mu^\mathrm{RI'} = \frac{1}{12 N_c} \gamma^\nu \left(g_{\mu\nu} - \frac{p_\mu p_\nu}{p^2} \right) \,,
    \label{eq:Pmu_RIprime}
\end{equation}
for which $Z_V=1$ when the amputation is performed with propagators renormalized with $\Zp$, since $P_\mu^\mathrm{RI'}$ obeys the relation
\begin{equation}
    \tr[P^{\mu,\mathrm{RI'}} \Lambda_\mu^V(p,p)] = \Zp \,.
    \label{eq:PmuRI_WI}
\end{equation}
From the physical point of view, we observe that the projector above attempts to restore the on-shell condition $F_1=\Sigma$ by projecting away the structure proportional to $p_\mu \slashed p$.

This interplay between the projector and the field renormalization is a well-known fact and the condition above practically defines the RI'/MOM scheme~\cite{Chetyrkin:1999pq,Gracey:2003yr}. The latter was introduced to overcome the practical difficulties in implementing $\Zg$ in Lattice QCD calculations, compared to the more easily accessible definition $\Zp$. 
In Refs.~\cite{Gracey:2003yr,Gracey:2010ci,Gorbahn:2022rgl}, $P_\mu^\mathrm{RI'}$ is derived by fixing the coefficients $c_i$ in
\begin{equation}
    P_\mu^\mathrm{RI'} = \frac{1}{12 N_c} \gamma^\nu T_{\mu\nu}^i(p,p) \, c_i \,,
\end{equation}
to be solutions of the linear system
\begin{equation}
    M^{ij} c_j = \{1,0\}^i \,,
\end{equation}
with the matrix
\begin{equation}
    M^{ij} = \frac{1}{3 N_c} T_{\alpha\mu}^i(p,p) T_{\beta\nu}^j(p,p) g^{\alpha\beta} g^{\mu\nu} \,.
\end{equation}
Our derivation, based on a simple linear combination of different variants of the projected Ward identity, confirms the results above.

\subsection{Non-exceptional kinematics}

Having established a framework parallel to the tensor decomposition of Refs. \cite{Gracey:2003yr,Gracey:2011fb,Gorbahn:2022rgl}, we now proceed to non-exceptional kinematics, a modern approach in momentum schemes due to the reduced sensitivity to non-perturbative effects at low energies \cite{Aoki:2007xm,Sturm:2009kb,Aoki:2005ga}.

The Ward identity for SMOM kinematics simplifies to
\begin{equation}
    q^\mu \Lambda_\mu^V(p,p') = i S^{-1}(q) = \slashed q \Zp \,,
    \label{eq:WI3}
\end{equation}
and when saturated with $\slashed q$,
\begin{equation}
    \frac{1}{4 N_c q^2} \tr[\slashed q q^\mu \Lambda_\mu^V(p,p')] = \Zp \,,
\end{equation}
it tells us that the projector $P_\mu^{\slashed q} = \frac{1}{4 N_c} \frac{q_\mu \slashed q}{q^2}$ leads to a corresponding factor $Z_V$ equal to unity~\cite{Sturm:2009kb}.
We conclude that the $\slashed q$ projector is perfectly suited to be promoted to the semi-leptonic case. 

To extend this scheme, we consider the following additional projections
\begin{align}
    \frac{1}{4 N_c p^2} \tr[\slashed p q^\mu \Lambda_\mu^V(p,p')] = & \frac12 \Zp \,, \\
    \frac{1}{4 N_c p^2} \tr[\slashed p' q^\mu \Lambda_\mu^V(p,p')] = & -\frac12 \Zp \,.
\end{align}
Following the idea of Ref.~\cite{Gorbahn:2022rgl}, we  build the appropriate linear combination of the two equations above to fix the r.h.s. to $\Zp$ and we find a new family of projectors for the vector bilinear in QCD
\begin{equation}
    P_\mu^{\mathrm{RI/SMOM}_y} = \frac{1}{2 N_c p^2} (y \slashed p + (y-1) \slashed p') q_\mu  \,,
    \label{eq:Pmu_RISMOM}
\end{equation}
satisfying the renormalized Ward identity for any value of $y$ and incorporating $P_\mu^{\slashed q}$ at $y=1/2$, the choice that was recognized as preserving the Ward identity in Ref.~\cite{Sturm:2009kb}.

Similarly to the RI/MOM case, we repeat the derivation using the approach based on the tensor decomposition in Eq.~\eqref{eq:tensor}, which is much richer due to the presence of the two momenta $p$ and $p'$. In fact, beyond a trivial extension of Eq.~\eqref{eq:Tmunu_MOM} to all possible pairs of momenta, the tensor basis also contains $\gamma^\mu \slashed p' \slashed p$, leading to~\cite{Gracey:2011fb}
\begin{equation}
    T_{\mu\nu}^i(p,p') = \left\{ g_{\mu\nu} \,, \frac{p_\mu p_\nu}{\mu^2}, \frac{p'_\mu p_\nu}{\mu^2}, \frac{p_\mu p'_\nu}{\mu^2}, \frac{p'_\mu p'_\nu}{\mu^2}, g_{\mu\nu} \frac{\slashed p' \slashed p}{\mu^2} \right\} \,.
\end{equation}

When inserted in Eq.~\eqref{eq:WI3}, the decomposition above translates in the following relation~\cite{Gracey:2011fb,Gracey:2010ci,Gorbahn:2022rgl}
\begin{equation}
    \slashed p[F_1 - \frac12 F_2 + \frac12 F_3] - \slashed p' [F_1 + \frac12 F_4 - \frac12 F_5 - F_6] = \slashed q \Zp \,.
\end{equation}
As noted in Ref.~\cite{Gorbahn:2022rgl}, this system with two linear equations offers the opportunity to define a family of projectors parametrized by $x$ in that reference. Therefore, the next natural step consists in considering
\begin{equation}
    F_1 + x\left(- \frac12 F_2 + \frac12 F_3\right) + (1-x)\left(\frac12 F_4 - \frac12 F_5 - F_6\right) = \Zp \,,
    \label{eq:SMOM_WI_FF}
\end{equation}
and in introducing the projector~\cite{Gorbahn:2022rgl}
\begin{equation}
    \frac{1}{4 N_c} \gamma^\nu T_{\mu\nu}^i(p,p') \, c_i \,.
\end{equation}
By adopting the same philosophy used for the RI/MOM scheme, the coefficients $c_i$ are found as solutions of the system (now adapted to the SMOM case from Eq.~\eqref{eq:SMOM_WI_FF})
\begin{equation}
    M^{ij} c_j = \left\{1, -\frac{x}{2},  \frac{x}{2}, \frac{1-x}{2}, -\frac{1-x}{2}, -(1-x) \right\}^i \,,
\end{equation}
with the matrix $M^{ij}$ for SMOM kinematics given by\footnote{Albeit with a different normalization, the matrix $M^{ij}$ was introduced in Appendix A of Ref.~\cite{Gracey:2010ci} using a slightly different tensor basis. We have checked that our findings hold in that case as well.}
\begin{equation}
    M^{ij} = \frac{1}{4 N_c} \tr\left[ \gamma^\alpha T_{\mu\alpha}^i \gamma^\beta T_{\nu\beta}^j \right] \, g^{\mu\nu} \,.
\end{equation}
After some algebra we find the solution vector
\begin{equation}
    c_i = \frac23 \left\{0, -(1+x), (1+x), -(x-2), (x-2), 0  \right\}_i \,,
\end{equation}
leading to the following compact form for the projector
\begin{equation}
\label{eq:solution-smom-proj}
    P_\mu^{\overline{\RI}/\SMOM_x} = 
    -\frac{1}{6 N_c \mu^2} q_\mu \left( (x+1) \slashed p + (x-2) \slashed p' \right) \,.
\end{equation}
We note that the solution above and the more compact Eq.~\eqref{eq:Pmu_RISMOM} are identical, up to the redefinition $y = \frac13 (x+1)$. Both derivations rely on simple observations from the Ward identity at the non-perturbative level.
Moreover we notice that the Dirac structures proportional to $F_1$ and $F_6$ are automatically excluded, consistently with the original derivation of the RI/SMOM scheme \cite{Sturm:2009kb}.

Finally, similarly to Ref.~\cite{Gorbahn:2022rgl} we also find that the projector
$\gamma_\mu / (16 N_c)$
leads to a $Z_V \neq 1$ and therefore it is not suitable for being promoted to the semi-leptonic case with SMOM kinematics, unless one modifies the wave-function renormalization condition \cite{Sturm:2009kb}.

\subsection{Fierz identities for the semi-leptonic operator}

Before concluding we show the equivalence between the double-trace projectors for the semi-leptonic operator derived above with the single-trace convention used in Ref.~\cite{Gorbahn:2022rgl}. To do so we rely on the Fierz identity
\begin{equation} 
    [P_R]^{\alpha\beta} [P_L]^{\gamma\delta} = 
    \frac{1}{2} [\gamma^\nu P_L]^{\alpha\delta} [\gamma_\nu P_R]^{\gamma\beta} \equiv 
    \frac{1}{2} \gamma^\nu P_L \tilde \otimes \gamma_\nu P_R \,,
    \label{eq:fierz-new}
\end{equation}
where the $\tilde \otimes$ denotes open tensor indices that result in single fermion traces when applied as a projector.

Factoring off the color index contraction and using the expression for $P_\mu^{\overline{\RI}/\SMOM_x}$ from Eq.~\eqref{eq:solution-smom-proj}, we rewrite our general form for the projector in Eq.~\eqref{eq:proj} for SMOM kinematics as
\begin{equation}
\begin{split}
    P^{\alpha\beta\gamma\delta} &=
   [P^\mu P_R]^{\alpha \beta}  [P_L \gamma_{\mu}]^{\gamma \delta} = \frac{1}{2} [P^\mu \gamma^{\nu} P_L \gamma_{\mu}]^{\alpha\delta} [\gamma_{\nu} P_R]^{\gamma\beta}  \\
   &=-\frac{1}{12N_c\mu^2} \left[(x+1)\slashed{p} + (x-2)\slashed{p}^\prime \right]\gamma^\nu\slashed q P_R\tilde{\otimes}\gamma_\nu P_R\,.
\end{split}
\end{equation}
After simple algebraic manipulations we find
\begin{equation}
\begin{split}
    P^{\alpha\beta\gamma\delta}=&\frac{1}{12N_c\mu^2}\left[\mu^2(x-2)\gamma^\mu P_R\tilde \otimes \gamma_\mu P_R \right.\\&\left.- 2(x+1)\slashed p P_R \tilde \otimes \slashed p P_R + 2(x+1)\slashed p^\prime P_R \tilde \otimes \slashed p P_R\right.\\
    &\left. +2(x-2)\slashed p^\prime P_R \tilde \otimes \slashed p^\prime P_R-2(x-2)\slashed p P_R \tilde \otimes \slashed p^\prime P_R\right.\\
    &\left.+(2x-1)\gamma^\mu\slashed p^\prime\slashed p P_R \tilde \otimes \gamma_\mu P_R\right],
\end{split}
\end{equation}
which matches the expression of the projector given in Eq.~(2.34) of Ref.~\cite{Gorbahn:2022rgl}. 
Similarly, plugging the expression for $P^\mu$ from Eq.~\eqref{eq:Pmu_RIprime} in Eq.~\eqref{eq:proj}, and making again use of the identity in Eq.~\eqref{eq:fierz-new}, yields the projector introduced in Eq.~(2.20) of Ref.~\cite{Gorbahn:2022rgl}.
This concludes our proof on the equivalence of the
two prescriptions.

\section{Conclusions}

With our work, by building on Ref.~\cite{Gorbahn:2022rgl}, we have introduced a family of projectors for the renormalization of the vector quark bilinear operator in QCD for RI/SMOM schemes which satisfy the Ward identity, in the sense that $Z_V=1$ to all orders in perturbation theory. 
This property becomes particularly convenient when the quark bilinear is paired with a lepton-neutrino pair to build a four-fermion EFT operator, relevant e.g. in hadronic $\tau$ decays. In this framework a matching with the full Standard Model is required and this is typically performed up to a fixed order in perturbation theory. More specifically, by considering isospin-breaking effects to first order the most relevant corrections are of $O(\alpha \alpha_s)$. As observed in Ref.~\cite{Gorbahn:2022rgl}, choosing schemes in which $Z_V=1$ leads to a smaller scale dependence of the Wilson coefficients and consequently to a better assessment of systematic errors due to missing higher orders, a very important aspect in high-precision calculations.

With our study, we have promoted the renormalization of the quark bilinear to the semi-leptonic operator, thereby obtaining a family of projectors that naturally satisfy the Ward identity in QCD. Our family elegantly incorporates $P_\mu^{\slashed q}$~\cite{Sturm:2009kb}, which remains a very natural choice also for the renormalization of $\O_{rs}$.

Thanks to Fierz identities we have proven the equivalence with Ref.~\cite{Gorbahn:2022rgl}, concluding that the Wilson coefficients calculated in that reference can be used also with the proposed double-trace form of the projectors derived in this work.

\section*{Acknowledgments}

PB, MB, CL and JP thank colleagues of the RBC and UKQCD collaborations for many valuable discussions and joint efforts over the years. 
PB was supported by US DOE Contract DESC0012704(BNL) and the Scientific Discovery through Advanced Computing (SciDAC) program LAB 22-2580.
This work is (partially) supported by ICSC - Centro Nazionale di Ricerca in High Performance Computing, Big Data
and Quantum Computing, funded by European Union – NextGenerationEU.
MG and SJ acknowledge partial support from the UK Science and Technology Facilities Council through grants ST/X000699/1 and ST/X000796/1, respectively.
The work of FM is supported by the Deutsche Forschungsgemeinschaft (DFG, German Research Foundation) under grant 396021762 — TRR 257 “Particle Physics Phenomenology after the Higgs Discovery”.

\bibliography{biblio}
\bibliographystyle{elsarticle-num} 

\end{document}